\documentclass[12pt]{article}
\usepackage{amsmath}
\usepackage{graphicx,color}
\usepackage{amssymb}
\usepackage[hypertex]{hyperref}
\usepackage{srcltx}

\newcommand{\be}{\begin{equation}}
\newcommand{\ba}{\begin{eqnarray}}
\newcommand{\ea}{\end{eqnarray}}
\newcommand{\ee}{\end{equation}}

\newcommand{\f}{\frac}
\newcommand{\s}{\sqrt}

\newcommand{\ti}{\tilde}
\newcommand{\ap}{\alpha}

\newcommand{\ddd}{\cdot\cdot\cdot}
\newcommand{\no}{\nonumber \\}
\newcommand{\la}{\langle}
\newcommand{\lb}{\rangle}
\newcommand{\bea}{\begin{eqnarray}}
\newcommand{\eea}{\end{eqnarray}}
\newcommand{\bes}{\begin{equation*}}
\newcommand{\beas}{\begin{eqnarray*}}
\newcommand{\eeas}{\end{eqnarray*}}
\newcommand{\bas}{\begin{array*}}
\newcommand{\eas}{\end{array*}}
\newcommand{\ees}{\end{equation*}}

\newcommand{\ep}{\epsilon}

\begin{document}

\begin{titlepage}
\thispagestyle{empty}

\begin{flushright}
YITP-14-103
\\
IPMU14-0360
\\
\end{flushright}


\begin{center}
\noindent{\textbf{Boundary States as Holographic Duals of Trivial Spacetimes}}\\
\vspace{2cm}
Masamichi Miyaji $^a$,
Shinsei Ryu $^{b}$,
Tadashi Takayanagi $^{a,c}$,
Xueda Wen $^b$
\vspace{1cm}

{\it
$^{a}$Yukawa Institute for Theoretical Physics,
Kyoto University, Kyoto 606-8502, Japan\\
$^{b}$Department of Physics, University of Illinois at Urbana-Champaign,
1110 West Green St, Urbana IL 61801, USA\\
$^{c}$Kavli Institute for the Physics and Mathematics of the Universe,\\
University of Tokyo, Kashiwa, Chiba 277-8582, Japan\\
}

\vskip 2em
\end{center}

\begin{abstract}
We study real-space quantum entanglement included in conformally invariant boundary states in
conformal field theories (CFTs).
First, we argue that boundary states essentially have no real-space entanglement
by computing the entanglement entropy when we bipartite the system into two spatial regions.
From the viewpoint of holography,
this shows that boundary states are dual to trivial spacetimes of zero spactime volume.
Next, we point out that
a continuous multiscale entanglement renormalization ansatz (cMERA)
for any CFTs
can be formulated
by employing a boundary state as its infrared unentangled state
with an appropriate regularization.
Exploiting this idea, we propose an approximation scheme of cMERA construction for general CFTs.
\end{abstract}

\end{titlepage}

\newpage


\section{Introduction}

Recently, there have been important progresses in understanding the structures of spacetimes in gravitational theories.
The holographic principle predicts that spacetimes in a gravitational theory are emergent
from more fundamental degrees of freedom \cite{Hol},
as explicitly realized in the AdS/CFT correspondence \cite{Maldacena}.
Moreover, latest studies strongly suggest that these fundamental degrees of freedom are given by quantum entanglement
in quantum many-body systems.
The holographic formula of the entanglement entropy offers one manifestation of
this connection between structures of spacetimes and quantum entanglement \cite{RT,Ra}.
Furthermore,
an explicit map between the structure of quantum entanglement and geometries of gravitational theories
can be established by using the tensor network construction of many-body quantum state,
in particular MERA (multi-scale entanglement renormalization ansatz) \cite{MERA}, as noted in \cite{Swingle}.
A continuum version of MERA, called cMERA is available \cite{cMERA} and its connection to
the AdS/CFT correspondence has been studied in \cite{NRT}.
Another interesting network construction was argued to be dual to gravity in \cite{Qi}.
Refer to, e.g., \cite{MS,MIH,SwC,HaMa,MNRT} for more progresses in this subject.

In this paper, we consider a holographic counterpart of the most fundamental spacetime: a trivial spacetime
with vanishing spacetime volume,
or in other words just a point-like universe.
As the holographic entanglement entropy shows, the area of spacetime is proportional to the entanglement entropy.
Therefore a trivial spacetime corresponds to a quantum state with no real-space entanglement.
From the viewpoint of its dual conformal field theory (CFT) in the setup of AdS/CFT,
such a quantum state with no entanglement is rather non-trivial,
since for typical states in CFTs,  such as the CFT vacuum states,
the real-space entanglement entropy scales with the size of the subregion of our interest and thus is far from vanishing
\cite{HLW,Cardy,RT}.
One of the main purposes of this paper is to point out that such an unentangled quantum state
can be obtained as the conformally invariant boundary states in CFTs.
The boundary states play an important role in CFTs
because they offers us general descriptions of boundary conformal field theories (BCFTs).
In 2d CFTs, the definition of boundary states, called the Cardy states \cite{CS},
is already interesting from a quantum information theoretic viewpoint
as their basic construction involves maximally entangled states of the holomorphic- and antiholomorphi-moving sectors
of the CFTs,
called the Ishibashi states \cite{Is}.
Identification of boundary states as a holographic dual to empty spacetimes
also helps us to find a construction of cMERA for generic CFTs;
Starting from such empty spacetime,
emergent spacetimes in the MERA or tensor network formalism
can be constructed by adding quantum entanglement to the corresponding quantum many-body state
at IR.

This paper is organized as follows:
In section \ref{Boundary States as Maximally Entangled States},
we give a quick overview of the construction of boundary states in CFTs and
interpret them as the maximally entangled states.
In section \ref{No Real-Space Entanglement in Boundary States},
we show that the boundary states have essentially no real-space entanglement.
In section \ref{cMERA for General CFTs and Boundary States},
we will start with a review of (c)MERA.
Later we will argue that a general construction of cMERA for CFTs can be obtained by using the boundary states.
In section \ref{Conclusions and Discussion},
we summarize our conclusions and discuss future problems.

\section{Boundary States as Maximally Entangled States}
\label{Boundary States as Maximally Entangled States}

In this section, we will review the definition of boundary states in CFTs,
focusing in particular on two-dimensional (2d) CFTs.
We will then explain an interpretation of boundary states as maximally entangled states.

\subsection{Basics of 2d CFTs}

One of the most salient features of 2d CFTs is the presence of an infinite dimensional symmetry, the Virasoro symmetry.
This symmetry is described by
two sets of Virasoro generators,
denoted by $\{L_n\}_{n\in \mathbb{Z}}$ and $\{\ti{L}_n\}_{n\in \mathbb{Z}}$
for the left- and right-moving sector, respectively.
They satisfy the Virasoro algebra:
\ba
&&[L_n,L_m]=(n-m)L_{n+m}+\f{c}{12}(n^3-n^2)\delta_{n+m,0},  \no
&&[\ti{L}_n,\ti{L}_m]=(n-m)\ti{L}_{n+m}+\f{c}{12}(n^3-n^2)\delta_{n+m,0},
\ea
where $c$ is the central charge of a given 2d CFT.
The vacuum state $|0\lb$ is defined by
\be
L_n|0\lb=\ti{L}_n|0\lb=0,\ \ \ (n\geq -1).
\ee
A primary state $|h,\bar{h}\lb$, the highest weight state of each irreducible representation of the Virasoro algebra, is defined by the condition
\ba
&& L_n|h,\bar{h}\lb={\ti{L}}_n|h,\bar{h}\lb=0 \ \ \ (n>0),\no
&& L_0|h,\bar{h}\lb=h|h,\bar{h}\lb,\ \ \ti{L}_0|h,\bar{h}\lb=\bar{h}|h,\bar{h}\lb. \label{vacvir}
\ea

Let us focus
on the left-moving sector
and consider states which are obtained
by acting the Virasoro generators on a primary state $|h\lb_L$,
so called the descendants of $|h\lb_L$:
\be
\ddd (L_{-n})^{k_n}(L_{-(n-1)})^{k_{n-1}}\ddd (L_{-1})^{k_1} |h\lb_L. \label{vstate}
\ee
We can choose an orthonormal basis
$|\vec{k},h\lb_L$, where $\vec{k}=(k_1,k_2,\ddd)$ is an infinite dimensional vector with components
all non-negative integers, such that ${}_L\la\vec{k},h|\vec{k'},h\lb_L=\delta_{\vec{k},\vec{k'}}$.
Especially, we would like to define such a basis by starting from the states of the form (\ref{vstate})
and deform into an orthonormal basis by taking linear combinations of states with the same level (i.e. eigenvalue of $L_0$).
Since the matrix, so called Gram matrix, given by the inner products of all states of the form (\ref{vstate}) is real-valued,
the states $|\vec{k},h\lb_{L}$ are obtained from linear combinations of states (\ref{vstate}) with real coefficients.

Then we introduce an infinite dimensional representation
$C_n$ of the Virasoro algebra acting on $|\vec{k},h\lb_L$:
\be
L_n|\vec{k},h\lb_L=\sum_{\vec{k'}}C^{\vec{k},\vec{k'}}_n|\vec{k'},h\lb_L.
\ee
Notice that all components $C^{\vec{k},\vec{k'}}_n$ are real as
the vectors $|\vec{k},h\lb_{L}$ are real-valued.
Also it is useful to note that the relation
\be
C^{\vec{k},\vec{k'}}_n=C^{\vec{k'},\vec{k}}_{-n}, \label{rfrl}
\ee
which follows from the relation
${}_{L}\la \vec{k},h|L_n|\vec{k'},h\lb_L={}_{L}\la \vec{k'},h|L_{-n}|\vec{k},h\lb_L$.

Note that we can equally define the right-moving states $|\vec{k},h\lb_R$ and repeat the same arguments.

\subsection{Constructing Boundary States as Maximally Entangled States}

Consider a 2d CFT defined on a cylinder described by the coordinate $(t,x)$, where
$t$ is the time coordinate and $x$ is the space coordinate with the periodicity $2\pi$.
 Now we would like to consider a quantum state at $t=0$
defined by the path-integral during the time period $-\ep<t<0$. At the time $t=-\ep$, we assume that there is a space-like boundary extending for all values of $x$. Finally we take the limit
$\ep\to 0$. The state defined in this way is called the boundary state $|B\lb$ and this depends on the choice of the boundary conditions. Especially we are interested in the boundary conditions which preserve a half of original Virasoro symmetries and such theories are called boundary conformal field theories (BCFTs). More explicitly, boundary states in BCFTs are defined by the condition
\be
(L_n-\ti{L}_{-n})|B\lb=0.\label{bndys}
\ee

We can find a simple solution to this condition as follows:
\be
|I_{h}\lb\equiv \sum_{\vec{k}}|\vec{k},h\lb_L\otimes |\vec{k},h\lb_R. \label{Ishi}
\ee
This state $|I_{h}\lb$ is called the Ishibashi state
for the primary state $|h\lb$
\cite{Is}.
This fact can be easily proved using (\ref{rfrl}) as follows:

\ba
\sum_{\vec{k}}L_n|\vec{k},h\lb_L\otimes |\vec{k},h\lb_R &=&\sum_{\vec{k},\vec{k'}}C^{\vec{k},\vec{k'}}_n|\vec{k'},h\lb_L \otimes |\vec{k},h\lb_R \no
&=&\sum_{\vec{k},\vec{k'}}|\vec{k'},h\lb_L \otimes C^{\vec{k'},\vec{k}}_{-n}|\vec{k},h\lb_R \no
&=&\sum_{\vec{k'}}|\vec{k}',h\lb_L\otimes \ti{L}_{-n}|\vec{k'},h\lb_R.
\ea
From the form (\ref{Ishi}),
it is clear that the Ishibashi states are maximally entangled states of the left- and right-moving sectors
of the CFT.

The boundary states which correspond to physical boundary conditions
are the so-called Cardy states \cite{CS}, denoted by $|C_\ap\lb$.
The Cardy states satisfy a consistency condition, so-called Cardy's condition or open-closed duality for partition functions on cylinders,
which requires that the same results should be obtained both in open and closed string calculations.
In general, the Cardy states (labeled by the index $\ap$) are given by special linear combinations of Ishibashi states:
\be
|C_\ap\lb=\sum_{h}B_{\ap,h}|I_h\lb,
\ee
where $h$ runs over all primaries.

The boundary states are singular in that their norms are infinite, and hence it is useful to introduce a regularization.
A simple way to regularize the norms is to perform the Euclidean time evolution by $e^{-\ep H}$,
where $H=L_0+\ti{L}_0-\f{c}{12}$.
Then the Ishibashi states look like
\be
\f{1}{\s{Z(\ep)}}\sum_{\vec{k}} e^{-\ep E(\vec{k})}|\vec{k},h\lb_L \otimes |\vec{k},h\lb_R,
\ee
where $E(\vec{k})$ is the eigenvalue of $H$ and we defined
$Z(\ep)=\sum_{\vec{k}} e^{-2\ep E(\vec{k})}$. If we trace out the right-moving part we find that the reduced density matrix looks like
\be
\rho(h,\ep)=\f{1}{Z(\ep)}\sum_{\vec{k}}e^{-2\ep E(\vec{k})}|\vec{k},h\lb_L { }_L\la \vec{k},h|.
\ee
This is regarded as the thermal distribution at temperature $T=1/2\ep$ of left-moving modes.
Therefore the entanglement entropy between the left- and right-moving sectors
coincides with the thermal entropy $S_{th}(h,\ep)$ of the either sector.
These facts were noted earlier in \cite{MNRT,Zayas} for free field theories.

If we turn to the Cardy states with the same $\ep$ regularization, the reduced density matrix after
tracing out the right-moving part is given by
\be
\f{1}{\sum_{h}(B_{\ap,h})^2}\sum_{h}(B_{\ap,h})^2\rho(h,\ep).
\ee

A few comments are in order.
First,
it is straightforward to generalize the above construction of the Ishibashi states to those
in the presence of current algebras.
While we omit details,
it may be useful to remember that for the $U(1)$ current algebra
we can have a compact expression of the Ishibashi states:
\be
|I_{h(\pm)}\lb=\exp\left(\pm\f{1}{\s{k}}\sum_{n=1}^\infty \f{1}{n}J_{-n}\ti{J}_{-n}\right)|h\lb, \label{cral}
\ee
where the ``level'' $k$ is a positive integer,
and $(J_n,\ti{J}_n)$ are the generators of left- and right-moving $U(1)$ current algebras
which satisfy  $[J_n,J_m]=[\ti{J}_n,\ti{J}_m]=kn\delta_{n+m}$.
This Ishibashi state satisfies the boundary conditions $(J_n\mp\ti{J}_{-n})|B\lb=0$.


Second,
it is straightforward to generalize the definition of boundary states in higher-dimensional CFTs,
though we have less analytical control compared with those in 2d CFTs.
For example, for a free massless scalar in $(d+1)$-dimensional flat spacetime,
boundary states which correspond to the Neumann $(+)$ and Dirichlet $(-)$ boundary conditions are explicitly given by
\be
|B(\pm)\lb=\exp\left[\pm\f{1}{2}\int d^d k\,  a^\dagger_{\vec{k}} a^\dagger_{-\vec{k}}\right]|0\lb,
\ee
where $a^\dagger_{\vec{k}}$ is the creation operator of a momentum $\vec{k}$ scalar particle (see also the appendix A in this paper).
This state again can be viewed as the maximally entangled state:
\be
|B(\pm)\lb=\prod_{\vec{k},k_1>0}\sum_{n_{\vec{k}}=0}^\infty (\pm 1)^{n_{\vec{k}}}|n_{\vec{k}}\lb_L\otimes |n_{\vec{k}}\lb_R,
\ee
where we decomposed the Hilbert space into those with $k_1>0$ (right-moving) and those with $k_1<0$
(left-moving), where $k_1$ is a component of the $d$-dimensional vector $\vec{k}$.
We also define the number states $|n_{\vec{k}}\lb_{L,R}$ as usual:
$|n_{\vec{k}}\lb_L=\f{1}{\s{n!}}(a^\dagger_{-\vec{k}})^{n_{\vec{k}}}|0\lb$ and $|n_{\vec{k}}\lb_R=\f{1}{\s{n!}}(a^\dagger_{\vec{k}})^{n_{\vec{k}}}|0\lb$.
We will not get into details of boundary states in more general CFTs in this paper.

\section{No Real-Space Entanglement in Boundary States}
\label{No Real-Space Entanglement in Boundary States}

We will now study the real-space entanglement of the quantum states defined by boundary states.
In the end, we will argue that boundary states have essentially no real-space entanglement,
though it is maximally entangled in terms of the decomposition of the Hilbert space
into the left and right-moving parts, as we noted in the previous section.

\subsection{Boundary States from Relevant Perturbations}

Let us consider a relevant perturbation to a given 2d CFT Hamiltonian $H$:
\be
H_M=H+M^{2-\Delta_O}\int dx\,  O(x), \label{tdd}
\ee
where $O(x)$ is a relevant operator with the conformal dimension $\Delta_O$
and $M$ represents the mass scale of this massive deformation.
We assume that under the renormalization group (RG) flow,
the theory flows into a trivial IR theory which has no propagating degrees of freedom.
We consider a quantum state $|\Omega_M\lb$ defined
as the ground state of the massive Hamiltonian $H_M$.
In particular, we take $M\equiv 1/\ep$ to be as large as the order of the UV cut off $\Lambda$,
in which case the infinitesimally small quantity $\epsilon \sim 1/\Lambda$ can be regarded as the lattice constant
in a discretized lattice regularization.
In this case, it is clear that all real-space entanglement vanishes for the scale much larger than $\ep$ since the Hamiltonian $H_M$, after a discretization,
is approximated by the one with infinitely many disentangled points with the lattice constant $\ep$.

\begin{figure}[ttt]
   \begin{center}
     \includegraphics[height=6cm]{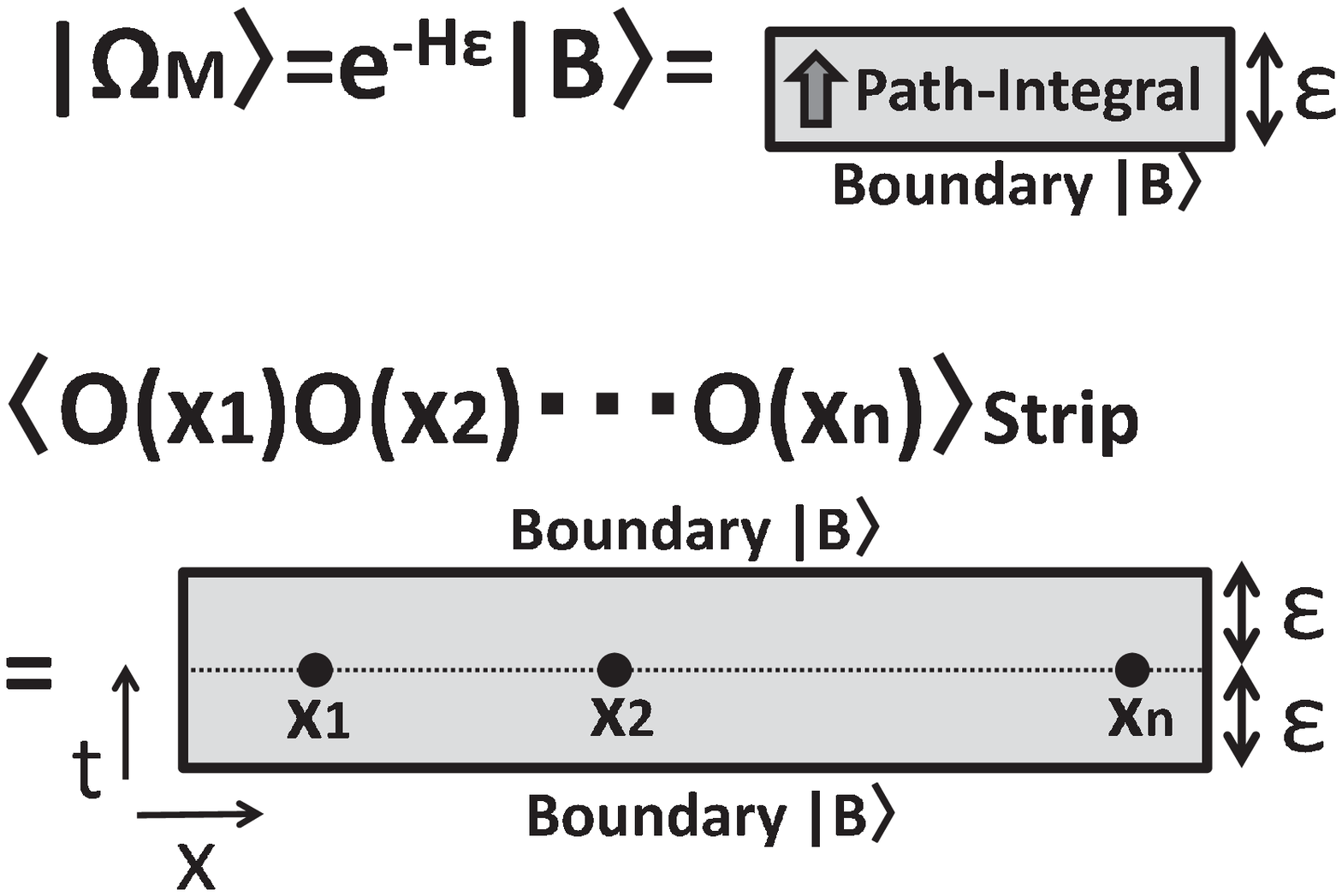}
   \end{center}
   \caption{Sketches of path-integral representation of the regularized unentangled state $|\Omega_M\lb$ and $n$ point functions $\la O(x_1)O(x_2)\ddd O(x_n)\lb_{Strip}$.}\label{statec}
\end{figure}

Actually, this procedure is equivalent to the (global) quantum quenches, where the massive deformation is suddenly turned off at a given (initial) time \cite{CaCaG}.
Therefore as argued in that context,
we expect that the quantum state $|\Omega_M\lb$ is given by a regularized boundary state:
\be
|\Omega_M\lb =\hat{F}(\ep)\cdot |B\lb,  \label{bs}
\ee
where $\hat{F}(\ep)$ is a regularization operation such that $\hat{F}(0)=1$.
As in \cite{CaCaG}, for example, we can approximately choose this to be, $\hat{F}(\ep)\propto e^{-\ep H}$.
Then the state (\ref{bs}) is described by a path-integral starting from the boundary for a (Euclidean) time period $\ep$ as in the upper picture in Fig.\ \ref{statec}.
This claim (\ref{bs}) can be understood because the infinitely large massive deformation
kills all dynamics, ending up with an empty theory and hence
the massive deformation introduces a boundary beyond which there is no degree of freedom.

As a simple example, let us consider a free massive Dirac fermion in (1+1)d \cite{Qi2012}:
\begin{align}
 H_M &= \int dx \left[
 -i v_F \psi^{\dag} \sigma_z \partial_x\psi + m \psi^{\dag}\sigma_x\psi
 \right],
\end{align}
where $\psi = (\psi_L, \psi_R)^T$ is a two-component fermion field,
$v_F$ is the Fermi velocity, and $m$ represents the mass term.
The ground state of this Hamiltonian is given by
\begin{align}
|\Omega_M\rangle
=
\exp\left[
\sum_{k>0}
\frac{ m}{\sqrt{ m^2 + (v_F k)^2 } + v_F k }
\left(
\psi^{\dag}_{L k} \psi^{\ }_{R k}
+
\psi^{\dag}_{R -k} \psi^{\ }_{L -k}
\right)
\right]
|{G}_L\rangle \otimes
|{G}_R\rangle
\end{align}
where
$\psi_{L,R k}$ is the Fourier component of $\psi_{L,R}(x)$, and
$|{G}_{L,R}\rangle$ is the Fock vacuum of the left- and right-moving sector in the massless limit, respectively.
As we take $m\to \infty$ ($m/(v_F k)\to \infty$),
$|\Omega_M\rangle$ reduces to the boundary states of the free massless fermion theory.

In summary,
we obtained from the above argument the first explanation that boundary states do not have real entanglement
by taking the cut off $\ep$ to zero.
Obviously we can generalize this result to higher dimensions.

Note also that the boundary states we constructed in the above massive deformation
should be the Cardy states $|C_\ap\lb$ as they describe physical boundaries.
For example if we take $O(x)=\phi(x)^2$ in a free scalar field theory (i.e., the $U(1)$ current algebra),
it is clear that we get the Cardy state corresponding to the Dirichet boundary condition.
It is a interesting future problem to find a general rule telling us which relevant perturbation in a bulk CFT leads to which boundary states.

\subsection{Correlation Functions}

It is also helpful to study correlation functions for quantum states given by the boundary states.
For this purpose consider the following $n$-point function of local operators at a particular time:
\be
\la O(x_1)O(x_2)\ddd O(x_n)\lb_{Strip}=
\f{\la B|e^{-\ep H}O(x_1)O(x_2)\ddd O(x_n)e^{-\ep H}|B\lb}{\la B|e^{-2\ep H}|B\lb},
\label{corw}
\ee
where $x_1,x_2,\ddd,x_n$ describe $n$ different spacial positions.
In the path-integral formalism with the regularization $\ep$,
this is evaluated as the path-integral over field configurations
on the strip shape spacetime defined by the region $0\leq t\leq 2\ep$
with the $n$ operators inserted at the time $t=\ep$,
as depicted in the lower picture of Fig.\ \ref{statec}.
There are two spacelike boundaries at $t=0$ and $t=2\ep$ which correspond to the boundary states.

In this setup it is obvious that in the limit $\ep\to 0$ with $x_i\neq x_j$ for all pairs $i\neq j$,
the correlation function (\ref{corw}) gets factorized into $n$ one point functions and does not depend on $x_i-x_j$.
This argument again shows that boundary states do not have real-space correlations.

\subsection{CFT Estimations of Entanglement Entropy}

We now give an explicit estimate of the entanglement entropy $S_A$ for,
as one of the simplest examples,
a free massless Dirac fermion theory when $A$ is an interval.
 We employ the results obtained in \cite{TaUg} where
the entanglement entropy under quantum quenches has been computed on a compact space.
We assume the free Dirac fermion theory on a cylinder $(t,x)$ with the periodicity $x\sim x+2\pi$.
We choose the boundary state $|B\lb$ to be either the Dirichlet or Neumann boundary condition
(in terms of the scalar obtained from the bosonization of the 2d massless Dirac fermion).

From the result in \cite{TaUg},
we find the following entanglement entropy $S_A$
for the interval $A$ of length $l$
for both the Neumann and Dirichlet boundary conditions:
\be
S_A=
\f{1}{6}\log
\f{|\theta_1\left(\f{l}{2\pi}|\f{2i\ep}{\pi}\right)|^2|\theta_1
\left(\f{\ep}{\pi i}|\f{2i\ep}{\pi}\right)|^2}
{\eta\left(\f{2i\ep}{\pi}\right)^6
\cdot |\theta_1\left(\f{\ep}{\pi i}-\f{l}{2\pi}|\f{2i\ep}{\pi}\right)|^2|\theta_1
\left(\f{\ep}{\pi i}+\f{l}{2\pi}|\f{2i\ep}{\pi}\right)|^2 \cdot a^2},
\label{SA }
\ee
where $a=1/\Lambda$ is the UV cut off (lattice spacing) of the free fermion theory.
Note that in the continuum limit $a\to 0$ with $\ep$ kept finite,
$S_A$ is logarithmically divergent,
$S_A\sim \f{1}{3}\log a^{-1}$, which is the well-known result in 2d CFTs \cite{HLW}
(remember the central charge is $c=1$ for the Dirac fermion).
If we assume our interested parameter region: $\ep\ll l$ and $\ep\ll 1$, the above entropy is simplified as
\be
S_A=\f{1}{3}\log \f{4\ep}{\pi a}.
\ee
Therefore we learned that if we take $\ep$ is of order of the UV cut off $a$, the logarithmically
divergent term in $S_A$ disappears and thus $S_A$ is finite, i.e.,  $S_A\sim O(1)$.
This shows that there is essentially no real-space entanglement in these boundary states.

It is also interesting to extend the above argument to the evaluation of $S_A$ for
the Cardy states in general 2d CFTs. Again we assume
that the subsystem $A$ is given by an interval of length $l$.
By using the replica trick,
we can represent $\mathrm{Tr}\, [\rho_A^n]$
as
the path-integral on $0\leq x\leq 2\pi$ and $0\leq t\leq 2\ep$
with two twisted vertex operators at the ends of the interval $A$ \cite{CaCaG}.
In the limit $l\gg \ep$, these two vertex operators are far apart and thus get decoupled.
The path-integral is then factorized into the two identical partition functions with single twisted operator.
Each of them can be conformally mapped to a disk partition function with the boundary condition specified by the Cardy state we consider.
Since the disk partition function is proportional to the $g$-function \cite{AL},
we obtain the following general result:
\be
S_A=\f{c}{3}\log\f{\ep}{a}+2\log g+c_1,  \label{eec}
\ee
where $c_1$ denotes an undetermined numerical constant which does not depend on the boundary condition.
Again this shows $S_A\sim O(1)$.
However, notice that the above result (\ref{eec}) is quantitatively correct only when $\ep\gg a$.
On the contrary, here we are interested in the case $\ep\sim a$ and therefore we cannot trust the precise value of the finite term in (\ref{eec}).
Nevertheless, at least we have succeeded to show that the logarithmically divergent term of the entanglement entropy
is absent for any boundary states and
this is enough to argue that boundary states essentially have no real-space entanglement.
Boundary states in CFTs with the continuum limit $\ep\to 0$
can still have finite amount of entanglement entropy, which may characterize their topological properties.
This finite entanglement entropy does not affect our main arguments in this paper.
We will explicitly evaluate this finite entanglement entropy in subsection 3.5 for a specific fermion model.

\subsection{Holographic Argument}

It is also useful to study the properties of boundary states from a holographic viewpoint.
For this purpose we can employ the AdS/BCFT correspondence introduced in \cite{TaBCFT}.
The gravity dual of the state $\rho\propto e^{-\ep H}|B\lb\la B|e^{-\ep H}$,
which is equivalent to a 2d CFT on a strip of width $2\ep$,
is given by a part of the Euclidean BTZ back hole. The metric of this (Euclidean) BTZ black hole is
\be
ds^2=R^2\left(\f{h(z)dt^2}{z^2}+\f{dz^2}{h(z)z^2}+\f{dx^2}{z^2}\right),\ \ \
h(z)\equiv 1-\f{\pi^2 z^2}{4\ep^2}. \label{btzm}
\ee
where $R$ is the AdS radius and the coordinates $(t,z,x)$ take values in the range
$-2\ep\leq t\leq 2\ep,\ 0<z\leq 2\ep/\pi,\ -\infty<x<\infty$. Note also that $t$ has the periodicity
$4\ep$.
Then let us divide this spacetime (\ref{btzm}) into two halves along a certain two-dimensional surface $Q$
which connects the two lines (extending in the $x$ direction) defined by $(t,z)=(0,0)$ and $(t,z)=(2\ep,0)$.
Our gravity dual of the CFT on the strip is given by the one which is surrounded both by
$Q$ and by the boundary half circle (defined by $z=0$ and $0<t<2\ep$).
In other words, the gravity dual is given by cutting out the BTZ black hole along the surface $Q$.

As argued in \cite{TaBCFT}, this surface $Q$ can be found by solving the
boundary equation of motion which involves the extrinsic curvature. The shape of $Q$ is related to
the $g$ function. The symmetric case where $Q$ is simply given by the middle half circle
$t=0$ and $t=2\ep$ is equivalent to the construction in \cite{HaMa}.

Now if we consider $n$ point functions (\ref{corw}) in the gravity dual of a 2d CFT on the strip, it is obvious there is no geodesic which connects $x_i$ and $x_j$ ($i\neq j$) because of the presence of the boundary $Q$.
Therefore there will be no correlations as long as $|x_i-x_j|\gg \ep$ in the approximation that the conformal dimensions of the local operators are very large.

We can also study the entanglement entropy. The holographic entanglement entropy is given by the area of minimal area surface divided by $4G_N$ \cite{RT}.
In the presence of spacetime boundary $Q$, the minimal surface can end on $Q$.
Therefore if the subsystem $A$ is given by an interval with the length $l$,
then the minimal surface is divided into two disconnected segments both of which connect the end points of $A$ to $Q$.
When $\ep\gg a$, we can approximate the metric (\ref{btzm}) by the pure AdS$_3$,
i.e., $h(z)=1$ and the surface $Q$ by two planes $t/z=const.$
and $(t-2\ep)/z=const$.
By employing the holographic calculation of the $g$-function in \cite{TaBCFT}, we finally find
\be
S_A\simeq \f{c}{3}\log \f{2\ep}{a}+2\log g,
\ee
which agrees with (\ref{eec}). Here $a$ is the UV cut off of the CFT and this corresponds to the UV
cut off $z>a$ in the gravity dual.

However, if we take $\ep$ to be of the same order as $a$, especially $\ep\leq \f{\pi}{2}a$, then the gravity dual itself becomes empty after we impose the UV cut off $z>a$. Therefore in such a situation
all correlation functions and entanglement entropy will vanish even with respect to finite contributions. These argument again shows that the gravity dual of boundary state $|B\lb$ (in the limit $\ep\to 0$) is the empty space and there is no real-space entanglement.

\subsection{Symmetry-Protected Topological Phases}

We have argued that the entanglement entropy of boundary states is given by a finite constant without any UV divergence. 
In the continuum limit analysis of CFTs, this constant seems to be non-universal in general. However, we can explicitly calculate its finite value in an explicit lattice model before we take the continuum limit. Indeed, this finite value of entanglement entropy of boundary states has an important implication in the context of symmetry-protected topological phases (SPT phases) in (1+1)d.
(For an overview of SPT phases, see \cite{TIReviews}).

In (1+1)d, there is no gapped phase with an intrinsic topological order, which in turn means
all gapped phases in (1+1)d can be continuously deformable (adiabatically connected).
to each other.
However, if some symmetry condition is imposed, two gapped phases may not be adiabatically connected,
i.e., they are separated by a quantum phase transition.
In particular, there may be a gapped phase which can be sharply (or ``topologically'') distinct from trivial gapped states
(such as an atomic insulator).
Such gapped phase is called a SPT phase.
Examples of SPT phases in (1+1)d include
topological insulators and superconductors, and the Haldane phase in integer spin chains.

It has been argued that different SPT phases (respecting the same symmetry conditions) can be
distinguished by their entanglement spectrum,
and, more precisely, by degeneracy in their entanglement spectrum
\cite{Pollmann}.
For example, in the Haldane phase, the entanglement spectrum shows a characteristic double degeneracy for each
entanglement eigenvalue, resulting in a $\log 2$ contribution to the entanglement entropy, whereas there is no such
degeneracy in trivial phases.
Thus, while there is no universal meaning in the constant entanglement entropy,
if there is some symmetry condition, it may limit the minimal possible entanglement entropy,
or the degeneracy of the entanglement spectrum.

To illustrate this point, we again go back to our free fermion example, but this case with a specific lattice regularization.
Consider a tight-binding Hamiltonian (the Su-Schrieffer-Heeger (SSH) model):
\begin{align}
H  &=
t_1
\sum_i
(c^{\dag}_{A i} c^{\ }_{B i}
+\mathrm{h.c.}
)
-
t_2
\sum_i
(
c^{\dag}_{B i} c^{\ }_{A i+1}
+
\mathrm{h.c.}
)
+
\mu_s
\sum_i
(c^{\dag}_{Ai} c^{\ }_{Ai} -c^{\dag}_{Bi} c^{\ }_{Bi}),
\end{align}
where $(c_{Ai}, c_{Bi})$ represents two-flavors of fermion annihilation operators defined at site $i$ on a 1d lattice,
and $t_1, t_2, \mu_s$ are real parameters.
We assume $t_1\ge 0$ and $t_2\ge 0$ henceforth.

When $\mu_s=0$, the Hamiltonian has chiral symmetry.
With this symmetry,
it is known that the phase with $t_1>t_2$ and $t_2>t_1$ are topologically distinct
and are always separated by a quantum critical point
\cite{Schnyder, Kitaev}.
(In the above model, there is a phase transition when $t_1=t_2$.)
Ground states in the phase $t_1>t_2$ are adiabatically connected to
an atomic insulator (a collection of decoupled lattice sites)
at $t_1>0$ and $t_2=0$,
i.e., the system is in the ``chiral topological insulator'' phase.
On the other hand,
ground states in the phase $t_2>t_1$ are topologically distinct from
topologically trivial, atomic insulators,
once chiral symmetry is imposed
\cite{RyuHatsugai}.
Near the critical point,
the SSH model reduces to the continuum Dirac Hamiltonian when expanded around $k=0$
with $t_1-t_2$ playing the role of the mass $m$.

%
%
The ground states of the SSH model
can be explicitly constructed as follows.
In momentum space, the Hamiltonian is written as
$H = \sum_k \Psi^{\dag}(k) \mathcal{H}(k) \Psi(k)$,
where
\begin{align}
\Psi(k)=
 \left[
 \begin{array}{c}
 c^{\ }_{k} \\
 d^{\ }_{k}
 \end{array}
 \right],
 \quad
\mathcal{H}(k)
=
\vec{R}(k) \cdot \vec{\sigma},
\quad
\vec{R}(k)
=
\left[
\begin{array}{c}
t_1 - t_2 \cos k \\
- \mu_s \\
t_2 \sin k
\end{array}
\right],
\end{align}
and $ -\pi \le k < \pi$;
We assume the theory is defined on a finite ring and the fermions obey a
suitable boundary conditions, e.g., antiperiodic boundary condition;
We have rotated the Pauli matrices as $(\sigma_x,\sigma_y, \sigma_z)\to (\sigma_x, \sigma_z, -\sigma_y)$
for convenience;
The field operator $\Psi(k)$ represents the fermionic annihilation operator in this rotated basis.
This Hamiltonian can be split split as
$H=H_L + H_R + H_{LR}$, where
$H_L$ and $H_R$ represent lattice analogues of the left- and right-moving parts of the continuum Dirac Hamiltonian
and are given by
\begin{align}
& H_L = t_2 \sum_k \sin (k) c^{\dag}_k c^{\ }_k,
 \quad
 H_R = -t_2 \sum_k \sin (k) d^{\dag}_k d^{\ }_k,
 \end{align}
 where as $H_{LR}$ plays the role of the mass and is given by
 \begin{align}
 H_{LR} =
 \sum_k (t_1 -t_2 \cos k + i\mu_s)
 c^{\dag}_k d^{\ }_k
 +
 \mathrm{h.c.}
\end{align}
By defining the Fock vacuum of $H_L$ and $H_R$ as
$|G_L\rangle$ and $|G_R\rangle$, respectively,
the (unnormalized) ground state of the total Hamiltonian is
\begin{align}
|\Omega\rangle
=
\exp \left[
\sum_{k>0}
\left(
\frac{ v_k }{u_k} c^{\dag}_k d^{\ }_k +
\frac{ u_{-k} }{v_{-k}} d^{\dag}_{-k} c^{\ }_{-k}
\right)
\right]
|G_L\rangle\otimes |G_R\rangle,
\end{align}
where
$u_k$ and $b_k$ are given in terms of $\vec{R}(k)$ by
\begin{align}
\left(
\begin{array}{c}
u_k \\
v_k
\end{array}
\right)
=
 \frac{R+ R_3}{\sqrt{2R(R+R_3)}}
\left(
\begin{array}{c}
R+R_3 \\
R_1 - i R_2
\end{array}
\right),
\quad
R(k)\equiv
|\vec{R}(k)|.
\end{align}
Setting $\mu_s=0$,
we are interested in the ground states deep inside the two phases (i.e., far away from the critical point):
\begin{align}
 |{G}\rangle
 &=
 \left\{
 \begin{array}{ll}
 \displaystyle
\exp \left[ - \sum_{k>0} \left( c^{\dag}_k d^{\ }_k + d^{\dag}_{-k} c_{-k} \right) \right] |G_L\rangle\otimes |G_R\rangle,
& t_1 >0, t_2=0
\\
\\
 \displaystyle
\exp \left[ + \sum_{k>0}
\frac{\cos k}{1+\sin k} \left( c^{\dag}_k d^{\ }_k + d^{\dag}_{-k} c_{-k}\right)
\right] |G_L\rangle\otimes |G_R\rangle,
&
t_1=0, t_2>0
 \end{array}
 \right.
 \label{BS SSH}
\end{align}

The ground state for $t_1>0, t_2=0$ represents an atomic insulator where
each site is completely isolated in the lattice model,
and hence $S_A=0$
for any bipartition of the 1d lattice.
Taking ``the continuum limit'', where we regard $c,d$ as $\psi_L, \psi_R$, respectively,
we reproduce the boundary state of the continuum massless Dirac theory.

The ground state for $t_1=0, t_2>0$ represents a topological insulator with ultra short correlation length.
By making the transition from the lattice fermion operator into the continuum Dirac fields as
$c\to \psi_L$ and $d\to \psi_R$,
{\it and} by taking the low energy limit $k\to 0$,
the ground state is again given by the boundary state.
While the correlation length of the ground state for $t_1=0, t_2>0$ is as short as the lattice constant,
the ground state has non-zero entanglement entropy.
Taking the subregion $A$ as an interval of arbitrary size,  we obtained a finite entanglement entropy $S_A= 2 \log 2$. 
(See Ref.\ \cite{RyuHatsugai} and the numerical calculation below.)

The finite entanglement comes from the UV physics, in that in the IR limit $k\to 0$
the ground states for topologically trivial and non-trivial phases show no difference
(except an overall minus sign in the exponential, which can be changed by simply reversing the sign of the
hopping parameter in the SSH model and hence is immaterial.)
This finite, constant contribution to the entropy, $2 \log 2$,
means that the entanglement spectrum defined for the subregion $A$ of arbitrary size
is at least doubly degenerate.
This double degeneracy is ``symmetry-protected''
and true for any ground state as far as the system is in the topological insulator phase.
Once we break the symmetry, i.e., chiral symmetry, by introducing non-zero $\mu_s$, say,
the symmetry protection is lost and the $l$-independent part of the entanglement entropy can take any value
depending on microscopic details.

\begin{figure}[ttt]
\begin{center}
(a)\includegraphics[width=0.4\columnwidth]{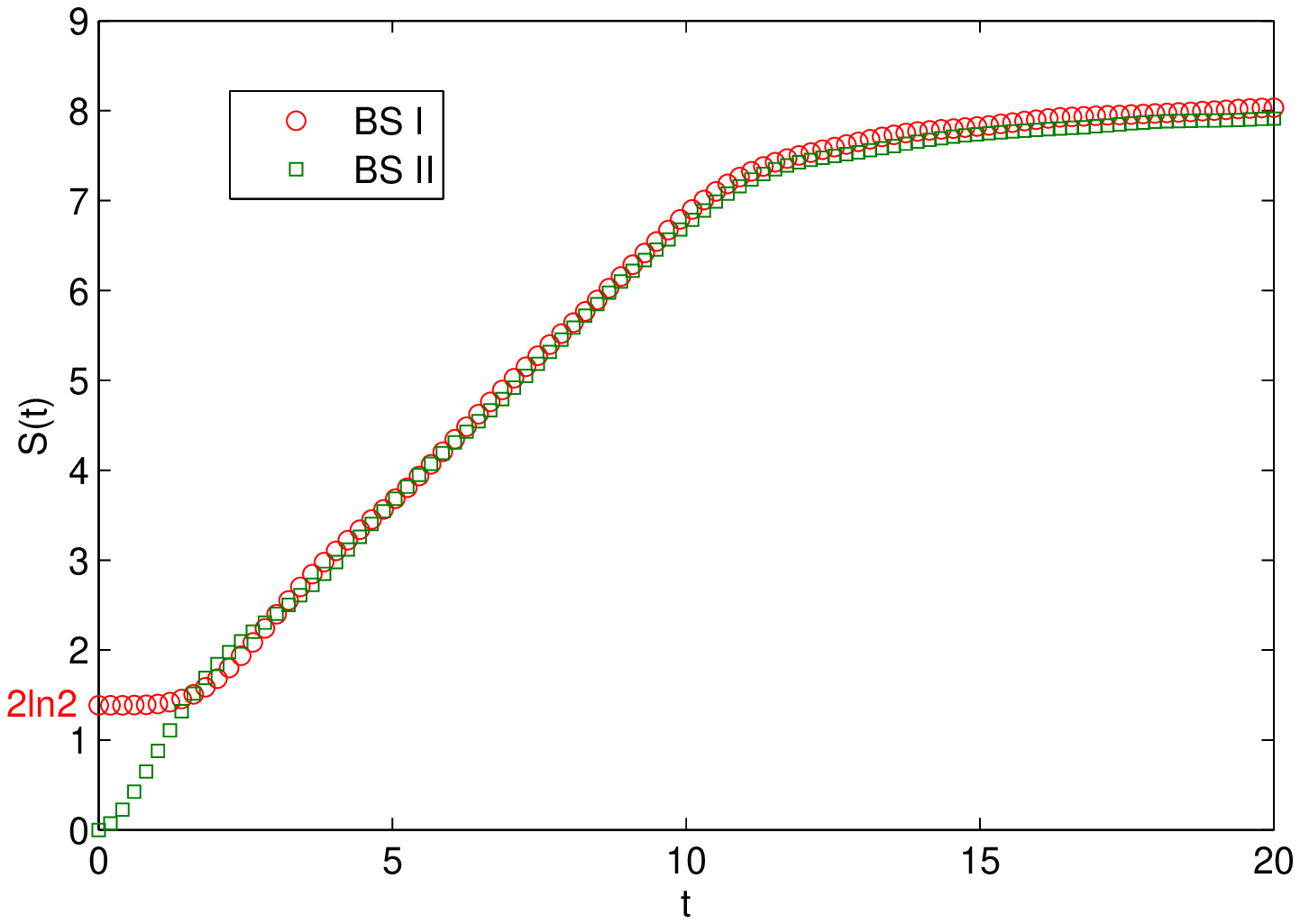}
(b)\includegraphics[width=0.4\columnwidth]{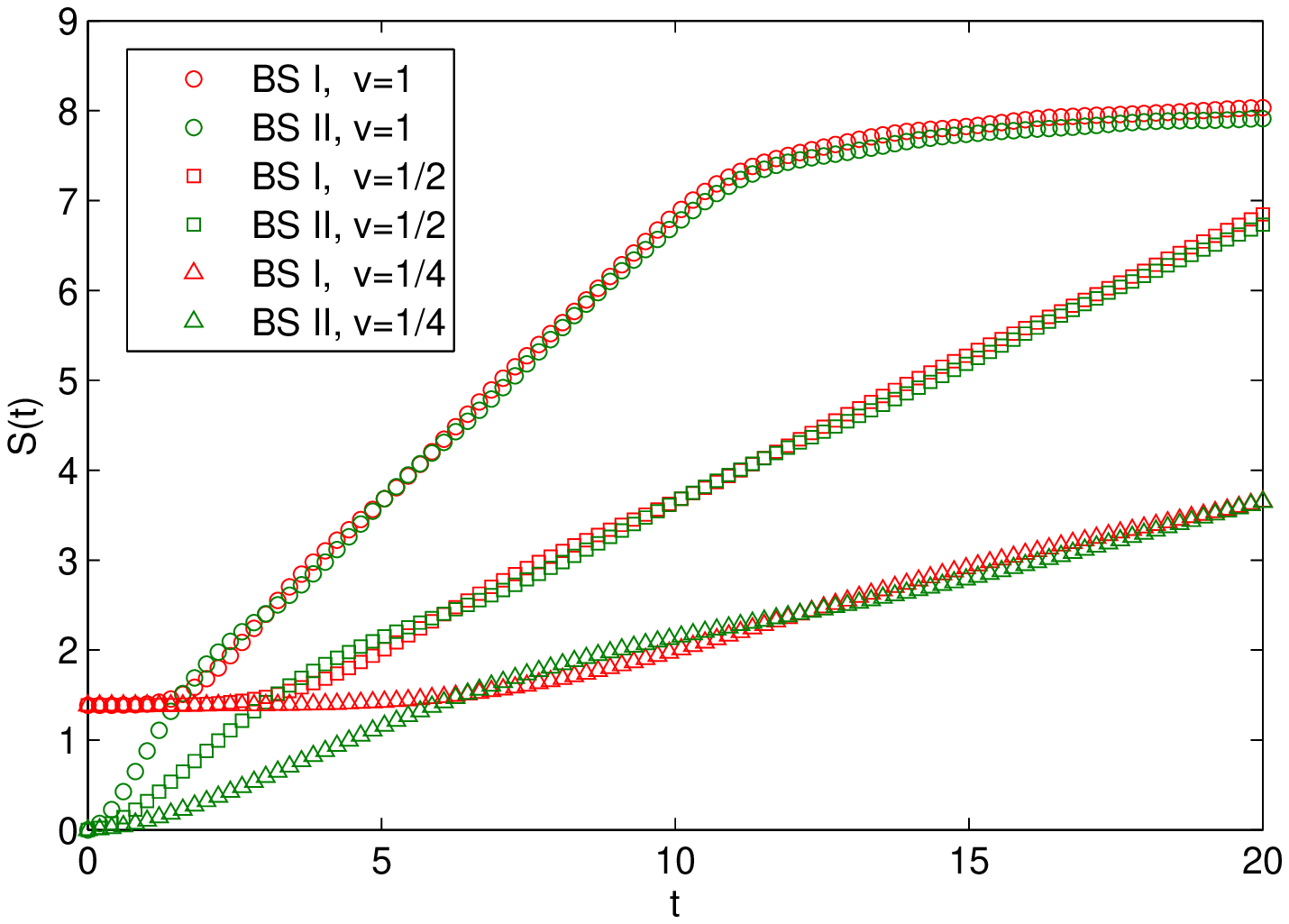}
\caption{
The numerically calculated time-evolution of the entanglement entropy, $S_A(t)$ of the SSH model,
after a global quench into the critical state for different boundary states
(denoted by ``BS I'' and ``BS II'', corresponding to
the case of $t_1=0, t_2>0$ and $t_1>0, t_2=0$ in  (\ref{BS SSH}), respectively).
We choose the size of the subregion $A$ $l=10$ and the total system size $L=200$
(both measured in the lattice constant).
(a) $S(t)$ as a function of time $t$.
(b) The velocity ($v$) dependence of  $S_A(t)$.
\label{fig:quench}
}
\end{center}
\end{figure}

In Fig.\ \ref{fig:quench},
a numerical calculation of
the time-evolution of the entanglement entropy $S_A$
after a quantum quench into the critical theory defined by $t_1=t_2$ and $m_2=0$ is presented,
which demonstrates, at $t=0$, $S_A=0$ for the topologically trivial ground state whereas
$S_A=2 \log 2$ for the topological insulator ground state.
On the other hand, for $t$ sufficiently larger than some threshold, $t_*$,
and for $t$ sufficiently smaller than the length of the region $A$, $l$,
the time evolution of the entanglement entropy is the one predicted from the CFT calculation
\cite{CaCaG, TaUg}. 
As mentioned above, the two possible values of $S_A$ at $t=0$ are quite robust,
as far as chiral symmetry is preserved.
For example, changing the velocity of the fermion mode 
(denoted by $v$ in Fig.\ \ref{fig:quench}, which is controlled by the hopping parameter $t_2$ in
the lattice model) 
changes the threshold value $t_*$,
while the value of $S_A(0)$ remains unaffected.

\section{cMERA for General CFTs and Boundary States}
\label{cMERA for General CFTs and Boundary States}

In this section,
by making use of our observation that boundary states in CFTs do not have real space entanglement,
we propose a formulation of cMERA for general CFTs.
We will first go through necessary ingredients of MERA
\cite{MERA}
and its continuous formulation called cMERA
\cite{cMERA}.
We will also give a brief overview on
the conjectured relation between MERA/cMERA and the AdS/CFT correspondence
\cite{Swingle,NRT}.
For a review on MERA and tensor-network methods to many-body systems in general,
refer to, e.g., \cite{Cir} for an excellent review.

\subsection{MERA and cMERA}

\paragraph{MERA}
MERA is a scheme for the real-space RG formulated in terms of wave functions.
This should be contrasted to the more familiar method of the Wilsonian RG,
where we consider the RG in terms of effective actions in momentum space.
Suppose we aim to find the ground state of a complicated Hamiltonian,
such as the one for a one-dimensional quantum spin system (spin chain).
Following the spirit of the real-space RG in classical statistical systems,
we consider to iteratively coarse-grain the spin chain by combining two quantum spins into one.
Let us define a non-positive integer $u$ which counts the iterative steps of this coarse-graining.
We describe the initial spin chain by $u=0$ and the first step of  coarse-graining is denoted by $u=-1$.
If we start with a spin chain with $N$ spins, after $|u|$ steps of coarse-graining the number of spins becomes
$N\cdot 2^{u}$.
In the end, it is reduced to a single spin after $\log_2 N$ steps.

This coarse-graining procedure alone, however,
does not give a good approximation of the correct ground state,
in particular when the quantum spin chain does not have a mass gap;
the amount of quantum entanglement included in the resulting wave function is much smaller than we wish,
in that its entanglement entropy $S_A$ has a finite upper bound.
On the other hand, $S_A$ in 2d CFTs increases logarithmically with respect to the size of $A$ \cite{HLW}.
To circumvent this problem in MERA,
a special type of tensor, called disentanglers, are introduced.
A disentangler is a unitary transformation which acts on nearest neighbor spins at each coarse-graining step,
and, as we go through the coarse-graining procedure, it
reduces bits of quantum entanglement of the original, highly entangled, ground state.

This MERA evolution, consisting of coarse-graining followed by disentanglers, can be run backward;
starting from a single spin,  we iteratively double the number of spins and, at each step, introduce quantum entanglement between adjacent spins
by a unitary transformation, which is the (dis)entanglers.
After many iterations, the MERA evolution reproduces the correct ground state.
These steps are the basic construction of MERA.
This formulation of $(1+1)$-dimensional MERA to higher dimensions
can be generalized in a straightforward way.

In \cite{Swingle}, it was conjectured that the MERA formulation of CFTs
is equivalent to the AdS/CFT correspondence by identifying the extra coordinate $u$
as the extra dimension.
The coordinate $u$ which counts MERA steps is expected to be related to
$z$, the standard radial coordinate of Poincar\'e AdS, as
$z=\ep\cdot 2^{-u}$, where
the metric of Poincar\'e AdS is given by
$
ds^2=R^2\left(dz^2+dx^2\right)/z^2
$,
and $\ep$ is the UV cut off in the AdS/CFT.
One important qualitative evidence for this conjecture
comes from
the fact that the estimation of the entanglement entropy in the MERA representation of quantum ground states
is given by minimizing the intersections between bulk surfaces and entangling bonds,
which looks analogous to the holographic entanglement entropy formula.

\paragraph{cMERA}

In order to apply MERA to continuum field theories,
the continuous MERA (cMERA) has been formulated
\cite{cMERA}.
We will consider cMERA for a $(d+1)$-dimensional QFT
and follow the convention of cMERA in \cite{NRT}.
In cMERA, we start from the unentangled state $|\Omega\lb$ (called the IR state)
and introduce the entanglement for each length scale
so that we can reproduce the correct state $|\Psi_{UV}\lb$ (the UV state),
which is typically a ground state of a given Hamiltonian.
This construction is naturally understood as a continuous limit of MERA.
Only apparent difference is that in cMERA
the dimension of the Hilbert space
(or the number of spins) does not change at each coarse-graining step.
In other words all procedures in cMERA such as coarse-graining and disentangling
are described by unitary transformations.
However, by adding dummy states at each coarse-graining step in MERA,
one can keep the total number of spins.
Refer to Appendix A in the present paper for the example of cMERA of free scalar fields.

We define a state $|\Psi(u)\lb$ parameterized by the scale $u$. This state $|\Psi(u)\lb$
is obtained by adding the entanglement for the momentum scale
$k\leq \Lambda e^{u}$ to the unentangled state $|\Omega\lb$.\footnote{
 We need be careful for this interpretation.
 This bound $k\leq \Lambda e^{u}$ is based on the viewpoint of UV state $u=0$.
 From the viewpoint of state $|\Psi(u)\lb$, the cut off is always $k\leq \Lambda$.
 Refer to our explanations later.
 }
The unentangled state (the IR state) $|\Omega\lb$ is defined as a state without any real-space entanglement.\footnote{
 In massless theories (CFTs) we need to be careful.
 The IR state $|\Omega\lb$ coincides with the unentangled state only
 if $k>\Lambda$.
 For $k<\Lambda$ we find $|\Omega\lb$ is equal to the CFT vacuum $|0\lb$.
 For more details see the example in Appendix A.2 and (\ref{psdf}).}
As will be argued in a later subsection, the IR state is closely related to the boundary states of CFTs.
We choose $\Lambda=1/\ep$ to be the original UV cut off scale,
where $\ep$ represents the lattice spacing.
When  $u=0$, the state $|\Psi(0)\lb$ includes all the entanglement and coincides with the UV state of our interest
(e.g. the ground state $|0\lb$ of a given QFT).
On the other hand, whtn $u=u_{IR}(=-\infty)$,
the state does not include any entanglement and is given by the IR state $|\Psi(u_{IR})\lb=|\Omega\lb$.

This procedure can explicitly be written as
\ba\label{unitaryTrnsf}
|\Psi(u)\lb = Pe^{-i\int^u_{u_{IR}}(K(s)+L)ds}|\Omega\lb,
\ea
where the symbol $P$ represents the path-ordering which puts all operators with smaller $u$ to the right.
The operator $K(s)$ describes (dis)entangler which introduces necessary entanglement along the MERA evolution
from $u=-\infty$ to $u=0$.
The entangler $K(s)$ acts only on the momentum modes with
$k\leq \Lambda$.
For this purpose we introduce the cut off function $\Gamma(k/\Lambda)$ so that $\Gamma(x)=1$ for $0<x<1$ and $\Gamma(x)=0$ in other cases \cite{cMERA}.
Schematically,  we have
\be
K(u)=\int d^d k\, \Gamma(k/\Lambda)\cdot g(u,k)\cdot O_k,
\ee
for any $u$, where $O_k$ is a suitable operator with the energy scale $k$.
We expect $\int d^d k\,O_k=\int d^dx\, O(x)$, where
$O(x)$ is a certain local operator in the CFT we consider.\footnote{
Note that $O_k$ is not a Fourier transformation of $O(x)$.
Refer to an explicit example in Appendix A.}
The function $g(u,k)$ depends on the theory and state of our interest and describes the strength of disentangling procedure.
The operator $L$ in (\ref{unitaryTrnsf}) is the non-relativistic scale transformation such that
\be
L|\Omega\lb=0.
\ee
In the MERA language, this operator $L$ corresponds to the corase-graining procedure.

\begin{figure}[ttt]
   \begin{center}
     \includegraphics[height=6cm]{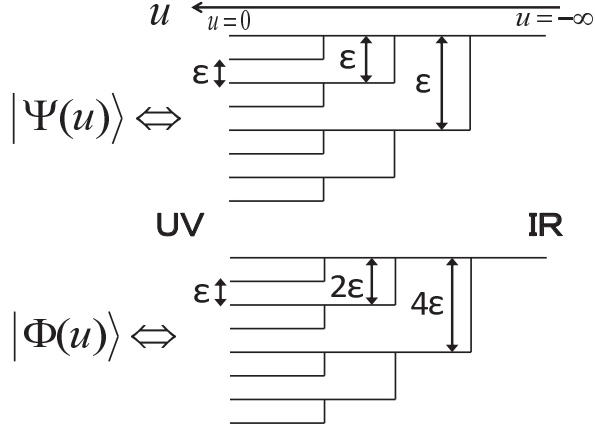}
   \end{center}
   \caption{Interpretations of two different cMERA states $|\Psi(u)\lb$ and $|\Phi(u)\lb$ in terms of MERA for discretized lattice models.
   In the former state, we always rescale the distance between two adjacent lattice points to be $\ep$,
   whereas in the latter state we do not rescale.}
   \label{MERAfig}
\end{figure}

To establish a connection between cMERA and MERA for discrete lattice models more closely,
we note, in the MERA picture, the effective lattice spacing is given by $\ep\cdot e^{-u}$ at the $(-u)$-th step.
Therefore the effective momentum cut off is given by $\Lambda e^{u}$.
However, in the description of cMERA (\ref{unitaryTrnsf}) using $|\Psi(u)\lb$,
we rescale the length scale at each step such that the momentum cut off is always given by $k\leq \Lambda$ for any values of $u$. In other words, we always rescale the effective lattice spacing to be $\ep$ for any $u$ in the description by $|\Psi(u)\lb$.
In the holographic dual language, we are looking at the co-moving coordinate $x$ in the AdS space.
This situation is depicted in the upper picture of Fig.\ref{MERAfig}.

When applying cMERA for a CFT vacuum state $|0\lb$,
there is a relativistic scale transformation $L'$ (i.e. the dilatation operator in the CFT).
It should be noted that $L$ and $L'$ are different because $L$ describes the spacial coarse-graining
or the space-like scale transformation.
The CFT vacuum $|0\lb$ satisfies $L'|0\lb=0$ but $L|0\lb$ is non-vanishing.
For a cMERA for a CFT vacuum, we can find the relation
\be
K(u)+L=L'
\quad
\mbox{for}
\quad
k\leq \Lambda.
\ee
This means that the IR state $|\Omega\lb$ is actually given by the vacuum $|0\lb$ for $k\leq \Lambda$,
which is a highly entangled state.
While this may sound puzzling at first sight,
this is simply an artifact of our rescaling procedure and nothing strange.
In our construction of $|\Psi(u)\lb$,
the non-relativistic scale transformation $L$ is performed at each step
and thus we disentangle degrees of freedom only for $k>\Lambda$ for \textit{any} value of $u$.
(On the other hand, if we consider
mass gapped theories rather than CFTs, then this seemingly singular behavior will not arise.
For more details of this issue, refer to (\ref{psdf}) in the explicit construction of cMERA for free scalars
in Appendix A.)

\paragraph{Reformulated cMERA}


In order to make a clearer connection to the MERA process,
it is useful to define a rescaled state $|\Phi(u)\lb$, which is defined by
\begin{align}
|\Phi(u)\lb &\equiv  e^{iuL}|\Psi(u)\lb
= Pe^{-i\int^u_{u_{IR}}\hat{K}(s)ds}|\Omega\lb.  \label{phfow}
\end{align}
Here $\hat{K}(s)$ is defined by the disentangler in the interaction picture
\be
\hat{K}(u)=e^{iuL}K(u)e^{-iuL}.
\ee
Therefore we have the representation
\be
\hat{K}(u)=\int d^dk\, \Gamma(ke^{-u}/\Lambda)\cdot g(u,ke^{-u})\cdot \ti{O}_k, \label{diskh}
\ee
where $\ti{O}_k$ is defined as $e^{iuL}O_k e^{-iuL}=e^{du}\ti{O}_{ke^u}$. This means that the effective lattice spacing at the scale $u$ behaves like $\ep\cdot e^{-u}$ which
agrees with the discrete MERA formulation.

To have a more concrete interpretation, let us assume a vacuum state for a free conformal field theory, where
$\hat{K}(u)$ is Gaussian:
\be
\hat{K}(u)\propto \int d^dk\,  \Gamma(ke^{-u}/\Lambda) b^\dagger_k b^\dagger_{-k},
\ee
where $b^\dagger_k$ is a creation operator of free field with momentum $k$ (see Appendix A).
This can be rewritten by defining a position space creation operator $b^\dagger_x\sim \int d^dk e^{ikx}b^\dagger_k$ as
\begin{align}
&
\hat{K}(u)\propto \int d^dx d^dy\, b^\dagger_x b^\dagger_y f(x-y),
\nonumber \\
\mbox{where}&
\quad
f(x-y)=\int d^dk\, e^{ik(x-y)}\Gamma(ke^{-u}/\Lambda).
\end{align}
We find that the function $f(x-y)$ takes non-trivial values only when $|x-y|\leq \ep\cdot e^{-u}$ is satisfied.\footnote{This observation is immediately understood by approximating the cutoff function by the gaussian $\Gamma(x)\sim e^{-x^2}$. However, the same conclusion can also be obtained for our original cutoff function $\Gamma(x)=1-\theta(|x|-1)$.}
This agrees with the discrete MERA formulation. Refer to the lower picture in Fig.\ref{MERAfig}.

\subsection{IR States for General CFTs}

In the above subsection we reviewed the idea of cMERA formulation.
However, if we are to construct cMERA for generic CFTs, we need to understand how to construct the IR state $|\Omega\lb$.
In this subsection we discuss a general construction of the IR state.
We focus on the formulation based on the state $|\Phi(u)\lb$ below. However we can always relate results from
$|\Phi(u)\lb$ to $|\Psi(u)\lb$ via the $L$ rescaling.
Our claims in this subsection can be confirmed explicitly in the free scalar example of appendix A.

To realize a disentangled IR state, as in (\ref{tdd}) we consider a relevant deformation of the original $d+1$ dimensional CFT Hamiltonian $H$:
\be
H_M=H+M^{(d+1)-\Delta_O}\int d^dx\, O(x), \label{relme}
\ee
where $O(x)$ is a relevant operator with the conformal dimension $\Delta_O$.
The infinitely large parameter $M$ represents the mass scale of this massive deformation, which is identified with the UV cut off (or lattice spacing) as $M=1/\ep$. We need to pick up an operator $O(x)$ which leads to a trivial IR fixed point under the RG flow. If we consider cMERA for a mass gapped QFT defined by a relevant perturbation of a CFT, we can choose $O(x)$ to be this relevant perturbation operator and our argument below still hold.

We now consider the ground state of $H_M$, denoted by $|\Omega_M\lb$,
and define it to be the IR state of cMERA, $|\Omega\lb$.
We propose that this is a general construction of IR states in any CFTs. Even though $H$ has the relativistic scale symmetry (dilatation) generated by $L'$ such that $[L',H]=iH$,
the mass deformed Hamiltonian $H_M$ does not have such a symmetry. Instead it is invariant under a non-relativistic scale transformation $L$ such that $[H_M,L]=0$. We can define $L$ such that the IR state $|\Omega\lb$, which is the ground state of $H_M$, is invariant under $L$ as $L|\Omega\lb=0$.

As we discussed in the argument below (\ref{tdd}), the IR state $|\Omega\lb$ defined as the ground state of $H_M$
is given by a regularized boundary state as in (\ref{bs}).
Especially in the limit $M\to\infty$ (or $\ep\to 0$), it should precisely approach to the boundary state  $|B\lb$.
Also it is useful to notice that the boundary state has the relativistic scale symmetry such that $L'|B\lb=0$.
Moreover, note that especially in 2d CFT cases, these boundary states should be regarded as the Cardy states instead of the Ishibashi states
because the infinitely large relevant perturbation leads to a physical boundary in the time direction.

\subsection{A General Simple Ansatz for cMERA}

Following the above argument,
one of the simplest ansatz of the IR state is
\be
|\Omega\lb={\cal N}e^{-\ep H}|B\lb, \label{statean}
\ee
where ${\cal N}$ is a normalization constant.
This is equivalent to the approximation used for quantum quenches in \cite{CaCaG}.
We now apply this approximation to cMERA for CFTs.
We will employ the rescaled formalism (\ref{phfow}).

For this purpose it is convenient to regard the UV cutoff $\Lambda$ in the disentangler as an artificial
cutoff for calculations which we can make much larger than the lattice spacing,
$\Lambda\gg M=1/\ep$:
By making $\Lambda$ arbitrarily large, we can make the MERA wave function for $k<1/\epsilon$
as close as possible to the true (ground) state.
(We are interested only in physics below the energy scale $M$.)
Finally we can determine the disentangler
$\hat{K}(u)$ from the relation (\ref{phfow}).
In order to get a simple analytical control,
we study the free scalar CFTs and the free fermion CFT in 2d below.

\subsubsection{Free Massless Scalar CFTs}

As the first example of a CFT, consider a free massless scalar field in any dimension.
We choose the disentangler in the following form (see also (\ref{Ksc}) in Appendix A)
\be
\hat{K}(u)=\f{i}{2}\int d^dk\, \Gamma(ke^{-u}/\Lambda)\left(g(u)a^{\dagger}_{k}a^{\dagger}_{-k}
-g^*(u)a_{k}a_{-k}\right). \label{KLsc}
\ee
Then the strength of disentangling procedure  at energy scale $\Lambda e^{u}$ is estimated by $|g(u)|$.
For our purpose we can assume that $g(u)$ is real-valued.
Then the relation (\ref{phfow}) with the anstaz (\ref{statean}) is written as
\be
2\int^0_{-\infty}du\, \Gamma(|k|e^{-u}/\Lambda)g(u)=\log (1+e^{-2\ep |k|})-\log (1-e^{-2\ep |k|}).
\ee
This relation is obtained from the Bogoliubov transformation description generally formulated in Section 3.2 of \cite{NRT}.
Assuming $\Lambda \ep\gg 1$, we can solve this as follows
\be
g(u)=\f{e^{\ti{u}}}{\sinh(2e^{\ti{u}})},
\ee
where we defined $\ti{u}$ by shifting $u$ as follows
\be
\ti{u}=\log(\Lambda\ep)+u.  \label{ure}
\ee
Note that in the physically interesting range $-\infty<\ti{u}<0$, the function $g(u)$ takes $O(1)$ non-vanishing values. This qualitatively tells us that the disentangling procedure continues for any $u$, which is consistent with the fact that we are considering a CFT.

\subsubsection{Free Massless Fermion CFT in Two Dimensions}

As another example, let us consider the free massless Dirac fermion in two dimensions.
We introduce the disentangler as (here we follow the notations in Ref.\ \cite{NRT}.)
\be
\hat{K}(u)=i\int dk \left[g_{k}(u)\psi^\dagger_1(k)\psi_2(k)+g^*_{k}(u)\psi_1(k)\psi^\dagger_2(k)\right],
\ee
where we defined
\be
g_k(u)=g(u)\Gamma(|k|e^{-u}/\Lambda)\f{ke^{-u}}{\Lambda}.
\ee
Again we can assume $g(u)$ is real-valued. We employ the approximated IR state (\ref{statean}) and solve the relation (\ref{phfow}), which leads to
\be
\cos\left[\f{ke^{-u}}{\Lambda}\int^0_{-\infty}du\Gamma(ke^{-u}/\Lambda)g(u)\right]=\f{1}{\s{1+e^{-4|k|\ep}}}.
\ee
Solving this equation, we obtain
\be
g(u)=\f{e^{\ti{u}}}{\cosh(2e^{\ti{u}})}+\arccos\left(\f{1}{\s{1+e^{-4e^{\ti{u}}}}}\right),
\ee
where $\ti{u}$ is defined as before (\ref{ure}).
Again, in the physically interesting range $-\infty<\ti{u}<0$, the function $g(u)$ takes $O(1)$ non-vanishing values.
This approximated solution improves the construction of cMERA for the free fermion in \cite{cMERA}, which essentially used the boundary state without regularizations as the IR state, since the latter solution can only be applied to the low energy region $k\ep\ll 1$.

\section{Conclusions and Discussion}
\label{Conclusions and Discussion}

In this paper,
we have studied quantum entanglement included in boundary states in CFTs.
Boundary states have been playing an important role
in boundary critical phenomena in condensed matter physics
and also in describing D-branes in string theory.
Our findings in this paper add more importance to boundary states from a quantum information theoretic viewpoint.


In particular,
we studied real-space entanglement of boundary states,
which can be measured by the entanglement entropy
when we spatially divide the system into two subsystems.
We gave several independent field theoretic as well as holographic arguments,
all of which show that there is essentially no real-space entanglement in boundary states.
While this may sound counterintuitive, given that
the left- and right-moving sectors are maximally entanglement in boundary states,
(almost) vanishing entanglement in boundary states may
be understood from entanglement monogamy,
one of the basic properties of quantum entanglement;
if two qubits A and B are maximally entangled, they cannot be entangled at all with a third qubit C.

From the holographic point of view,
the above result suggests that boundary states are dual to trivial spacetimes of zero spacetime volume,
as follows from the holographic entanglement entropy.
(We should however keep in mind that in such a situation, we will have substantial quantum gravity effects.)

The vanishing entanglement in boundary states
also has an implication and application in
MERA
which we believe captures fundamental aspects of the mechanism of the AdS/CFT correspondence.
In the MERA construction,
starting from a unentangled state (IR state),
quantum entanglement is added at each length scale from IR to UV
to reproduce a highly entangled state
of our interest, such as the ground state of a complex many-body Hamiltonian.
This procedure has a holographic interpretation in which,
starting from a trivial spacetime, gradually adding pieces of spacetime
eventually leads to a gravitational theory in an extended spacetime.
By using a continuum limit of MERA (cMERA),
we identified the IR unentangled state with (regularized) boundary states.
Our formulation allows a construction of cMERA which is universal in that it can be applicable to generic CFTs.
We confirmed that this new construction leads to sensible results for free massless scalar field theories.

For a given CFT, there are various choices of boundary states.
In our formulation they correspond to various choices of the relevant deformation (\ref{relme}).
This means, in fact, that the choice of the IR state $|\Omega\lb$ is not unique.
It is an important future problem to find a general rule as to which relevant deformation leads to which boundary state.
Related to this non-uniqueness,
it would be interesting to speculating upon
formulating the cMERA for the canonical examples of the AdS/CFT,
so that it is dual to taking slices in the AdS spacetime.
We then need to preserve the R-symmetry, which is dual to the symmetry of internal manifolds such as S$^5$ in AdS$\times$S$^5$.
In the setup of a global AdS in dimensions higher than three,
it is easy to identify the relevant deformation in the dual CFT on the compact spacetime with the conformal mass terms of the CFT $\sim \int R\phi^2$,
which manifestly preserves the R-symmetry.
It is natural that for the Poincare AdS setup,
we can still employ the same form of mass terms to define the cMERA for a CFT in a non-compact space. It will be important to understand a systematic relation between the choices of boundary states and their gravity duals.

In addition, in order to relate the cMERA formulation to the AdS/CFT,
we expect that taking the large-$N$ or large central charge limit of CFTs is necessary.
It is therefore an important future task to study the behavior of boundary states in this limit.
Roughly speaking, we can regard the unitary transformation of
a vacuum state (\ref{vacvir}) into a boundary state (\ref{bndys}) as a generalized version of the Bogoliubov transformation.
This is manifest in the $U(1)$ current example as in (\ref{cral}).
It is also curious to note that the Gaussian form of (\ref{cral}) is analogous to the double trace deformations which appear
in the analysis of the holographic RG \cite{HP}.

\section*{Acknowledgements}

We would like to thank
Gil Young Cho,
Pedro Lopes,
Shunji Matsuura,
Masahiro Nozaki,
Tokiro Numasawa,
Hirosi Ooguri, Xiaoliang Qi, Mark Van Raamsdonk, Mukund Rangamani and Masaaki Shigemori for useful discussions.
TT is very grateful to the workshop `Quantum Information Physics (YQIP2014)' at Yukawa Institute for Theoretical Physics, Kyoto University and the workshop `Quantum Information in Quantum Gravity' at University of British Columbia in Vancouver for giving opportunities to present a part of this work in very stimulating atmospheres.
TT is supported by JSPS Grant-in-Aid for Scientific
Research (B) No.25287058 and JSPS Grant-in-Aid for Challenging
Exploratory Research No.24654057. TT is also
supported by World Premier International
Research Center Initiative (WPI Initiative) from the Japan Ministry
of Education, Culture, Sports, Science and Technology (MEXT).
SR is supported by Alfred P. Sloan foundation.

\appendix

 \section{Example: cMERA for Free Scalar Field Theory}

Consider the free scalar field theory in $d+1$ dimensions (with mass $m$),
which is one of the simplest examples of cMERA \cite{cMERA,NRT}.
The time and space coordinates are denoted by $t$ and $x$.
The energy and the momentum are written as $\ep$ and $k$, respectively,
and the dispersion relation is given by $\ep_k=\s{k^2+m^2}$. The Hamiltonian of this theory is defined by
\be
H=\f{1}{2}\int d^d k\, \left[\pi(k)\pi(-k)+\ep^2_k\phi(k)\phi(-k)\right].
\ee
The creation and annihilation operators,
$a^{\dag}_{k}$ and $a_{k}$,
can be introduced
as follows
\ba
\phi(k)=\f{a_k+a^\dagger_{-k}}{\s{2\ep_k}},
\quad
\pi(k)=\s{2\ep_k}\left(\f{a_k-a^\dagger_{-k}}{2i}\right).
\ea
They satisfy the canonical commutation relation $[a_{k},a^{\dagger}_{k'}]=\delta^d(k-k')$.

\subsection{Construction of cMERA}

In the IR limit, the Hamiltonian becomes infinitely many copies of decoupled
harmonic oscillators located at each lattice point.
We can define an unentangled state
$|\Omega_{M_m}\lb$ as the ground state for this harmonic oscillators,
i.e., $a_x|\Omega_{M_m}\lb=0$.
Here $a_x=\s{M_m}\phi(x)+{i}\pi(x)/\sqrt{M_m}$
is the annihilation operator in real-space,
and $M_m$ is the energy scale below which the disentanglement takes place.
In other words, we assume the system is discretized into a lattice with lattice constant $\sim 1/M_m$.
In momentum space, this condition is equivalent to
\be
(\ap_k a_k+\beta_k a^{\dagger}_{-k})|\Omega_{M_m}\lb=0, \label{irs}
\ee
where
\ba
&&\ap_k=\f{1}{2}\left(\s{\f{M_m}{\ep_k}}+\s{\f{\ep_k}{M_m}}\right),\ \
\beta_k=\f{1}{2}\left(\s{\f{M_m}{\ep_k}}-\s{\f{\ep_k}{M_m}}\right),\no
&& M_m=\s{\Lambda^2+m^2}.
\ea

Since we are dealing with a non-interacting theory,
we assume the disentangler  $\hat{K}$ (\ref{diskh})
is ``Gaussian'' and takes the following form
\be
\hat{K}(u)=\f{i}{2}\int d^dk\, \Gamma(ke^{-u}/\Lambda)\left(g(u)a^{\dagger}_{k}a^{\dagger}_{-k}
-g^*(u)a_{k}a_{-k}\right), \label{Ksc}
\ee
where $\Gamma(x)$ is a cut off function such that $\Gamma(x)=1$ when $x\leq 1$ and $\Gamma(x)=0$ for $x>1$.
Indeed, by requiring that the UV state to be the CFT vacuum
\be
|\Phi(0)\lb=|0\lb,
\ee
(or equally $|\Psi(0)\lb=|0\lb$), we find the disentangler which reproduces the exact ground state as \cite{cMERA}
\be
g(u)=g^*(u)=\f{1}{2}\cdot \f{e^{2u}}{e^{2u}+m^2/\Lambda^2}. \label{scge}
\ee

\subsection{Scale Transformation in Massless Scalar}

The non-relativistic and relativistic scale transformation in the free boson theory are defined as follows
\cite{cMERA}
\ba
&& L=-\f{1}{2}\int d^dx \left[\pi(x)\vec{x}\cdot\vec{\nabla}_x\phi(x)+\vec{x}\cdot\vec{\nabla}_x\phi(x)\pi(x)+\f{d}{2}\phi(x)\pi(x)
+\f{d}{2}\pi(x)\phi(x)\right],\no
&& L'=-\f{1}{2}\int d^d x \left[\pi(x)\vec{x}\cdot\vec{\nabla}_x\phi(x)+\vec{x}\cdot\vec{\nabla}_x\phi(x)\pi(x)+\f{d-1}{2}\phi(x)\pi(x)
+\f{d-1}{2}\pi(x)\phi(x)\right].\nonumber
\ea
Note that the latter is the same as the dilatation operator in the CFT.
It is easy to see that the massive Hamiltonian $H_M=\f{1}{2}\int d^dx [\pi(x)^2+M^2 \phi(x)^2]$ and the
IR state $|\Omega\lb$ is invariant under the transformation by $L$.
Their actions on the creation/annihilation operators are given by
\ba
&& e^{-iuL'}a_k e^{iuL'} =e^{-\f{d}{2}u}a_{ke^{-u}},\no
&& e^{-iuL}a_k e^{iuL}=e^{-\f{d}{2}u}(\cosh(u/2)a_{ke^{-u}}+\sinh(u/2)a^{\dagger}_{-ke^{-u}}).
\ea

From the analysis of \cite{NRT}, we find
\ba
\left\{
\begin{array}{ll}
\left[\cosh\f{u}{2}a_k-\sinh\f{u}{2}a^\dagger_{-k}\right]|\Phi(u)\lb=0, & \mbox{for}\ k\leq \Lambda e^u,
\\
\\
\left[\ap_k\cdot a_k + \beta_k \cdot a^\dagger_{-k}\right]|\Phi(u)\lb=0, & \mbox{for}\ k\geq \Lambda e^u.
\end{array}
\right.
\label{mlphi}
\ea
By performing the $L$ transformation, we can rewrite these into
\ba
\left\{
\begin{array}{ll}
a_{ke^{-u}}|\Psi(u)\lb=0, & \mbox{for}\ k\leq \Lambda e^u,
\\
\\
\left[\ap_{ke^{-u}}a_{ke^{-u}} + \beta_{ke^{-u}} a^\dagger_{-ke^{-u}}\right]|\Psi(u)\lb=0, & \mbox{for}\ k\geq \Lambda e^u.
\label{mlpsi}
\end{array}
\right.
\ea
Therefore, in the massless scalar field example, we can explicitly find (when we decompose the state into momentum basis)
\ba
|\Phi(u)\lb &\propto&\prod_{k<\Lambda e^u}
\exp\left(\tanh\f{u}{2} a^\dagger_{k}a^\dagger_{-k}\right)|0\lb \cdot
\prod_{k>\Lambda e^u}|\Omega_\Lambda\lb , \label{pppf}
\ea
and
\ba
|\Psi(u)\lb &\propto &\prod_{k<\Lambda} |0\lb \cdot
\prod_{k>\Lambda} |\Omega_\Lambda\lb, \label{psdf}
\ea
where we recall that $|\Omega_{\Lambda}\rangle$ is defined in (\ref{irs}).

It is important to note that the cMERA state $|\Psi(u)\lb$ in the CFT is given by the unentangled state
$|\Omega_{\Lambda}\lb$ only for $k>\Lambda$. However, for $k< \Lambda$ we find that
$|\Psi(u)\lb$ is given by the CFT vacuum $|0\lb$. This look at first confusing if we take the IR limit $u=u_{IR}\to -\infty$ because $|\Psi(u_{IR})\lb$ should be the unentangled state. However, in the definition of
$|\Psi(u)\lb$ we always perform the non-relativistic scale transformation $L$ under the coarse-graining procedure and thus we only have the disentangled state for the short wave length $k>\Lambda$ for any $u$ (refer to Fig.\ref{MERAfig}). Moreover,
we should have $|\Psi(u)\lb=|0\lb$ for $k< \Lambda$ so that $K(u)+L$ becomes identical to the relativistic scale transformation $L'$. As we will see in the next subsection, we will not have this issue for the massive scalar example.

On the other hand, if we take the IR limit $u=u_{IR}\to -\infty$, then we find
$|\Phi(u_{IR})\lb=|\Omega_\Lambda\lb$ for any $k$. This is because we did not perform any scale transformation to define $|\Phi(u)\lb$. Notice also these facts are consistent with the fact that $|\Psi(0)\lb=e^{iuL'}|\Psi(u)\lb$
for $k\leq \Lambda$, where $L'$ is the CFT scale transformation.

Finally, note that from the above definition of $L'$, it is clear that $L'$ annihilates
both the vacuum $|0\lb$ and boundary state $|B\lb$:
\be
L'|0\lb=L'|B\lb=0,
\ee
where the boundary state (for Dirichlet boundary condition) is given up to the overall normalization constant by:
\ba
|B\lb &\propto&\prod_{k}
\exp\left(-a^\dagger_{k}a^\dagger_{-k}\right)|0\lb. \label{scbsg}
\ea

\subsection{Massive Scalar Case}

Here we generalize the massless results in the previous subsection into those in the massive scalar case.
The scale transformations are given by
\begin{align}
e^{-iuL}a_ke^{iuL}&=e^{-\f{du}{2}}\left[\s{\f{\ep_k}{\ep_{ke^{-u}}}}(a_{ke^{-u}}
+a^{\dagger}_{-ke^{-u}})+\s{\f{\ep_{ke^{-u}}}{\ep_{k}}}(a_{ke^{-u}}
-a^{\dagger}_{-ke^{-u}})\right],
\no
 e^{-iuL'}a_ke^{iuL'}&=
 e^{-\f{du}{2}}\left[
e^{-\frac{u}{2}}\s{\f{\ep_k}{\ep_{ke^{-u}}}}(a_{ke^{-u}}
+a^{\dagger}_{-ke^{-u}})
+e^{\frac{u}{2}}\s{\f{\ep_{ke^{-u}}}{\ep_{k}}}(a_{ke^{-u}}
-a^{\dagger}_{-ke^{-u}})\right]. \no
\end{align}
The state $|\Phi(u)\lb$ satisfies
\ba
\left\{
\begin{array}{l}
\left[\left(\s{\ep_{\Lambda}/\ep_{\Lambda e^{u}}}+\s{\ep_{\Lambda e^{u}}/\ep_{\Lambda }}\right) a_k-\left(
\s{\ep_{\Lambda}/\ep_{\Lambda e^{u}}}-\s{\ep_{\Lambda e^{u}}/\ep_{\Lambda }}\right)
a^\dagger_{-k}\right]|\Phi(u)\lb=0
\\
\qquad \qquad \mbox{for}\ k\leq \Lambda e^u,
\\
\left[\ap_k\cdot a_k + \beta_k \cdot a^\dagger_{-k}\right]|\Phi(u)\lb=0
\qquad \mbox{for}\ k\geq \Lambda e^u.
\end{array}
\right.
\ea
On the other hand, the state $|\Psi(u)\lb$ is found to be
\ba
\left\{
\begin{array}{l}
\left[\left(\s{\f{\ep_{\Lambda}\ep_k}{\ep_{\Lambda e^{u}}\ep_{ke^{-u}}}}+\s{\f{\ep_{\Lambda e^{u}}\ep_{k e^{-u}}}{\ep_{\Lambda }\ep_{k}}}\right)a_{ke^{-u}}-\left(
\s{\f{\ep_{\Lambda}\ep_k}{\ep_{\Lambda e^{u}}\ep_{k e^{-u}}}}
-\s{\f{\ep_{\Lambda e^{u}}\ep_{k e^{-u}}}{\ep_{\Lambda }\ep_k}}\right)a^\dagger_{-ke^{-u}}\right]|\Psi(u)\lb=0 \no
\qquad \mbox{for}\ k\leq \Lambda e^u,
\\
\left[\ap_{k e^{-u}}\cdot a_{ke^{-u}} + \beta_{ke^{-u}} \cdot a^\dagger_{-ke^{-u}}\right]|\Psi(u)\lb=0
\qquad \mbox{for}\ k\geq \Lambda e^u.
\end{array}
\right.
\ea
Thus we find for $k\leq \Lambda$
\ba
&& |\Phi(0)\lb=|0\lb,\ \ \ |\Phi(u_{IR})\lb=|\Omega_{M_m}\lb, \no
&& |\Psi(0)\lb=|0\lb,\ \ \ |\Psi(u_{IR})\lb=|\Omega_{M_m}\lb,
\ea
where we employed the properties in the IR limit $u=u_{IR}\to -\infty$:
\be
\Lambda\cdot e^{u_{IR}}=0,\ \ \ \ \ k\cdot e^{u_{IR}}=0.
\ee
Note also that for $k>\Lambda e^u$, we simply get $|\Phi(u)\lb=|\Omega_{M_m}\lb$
for any $u$.
And for $k>\Lambda$ we have $|\Psi(u)\lb=|\Omega_{M_m}\lb$ for any $u$.
In this way, for massive theories, for any momentum we obtain $|\Psi(u)\lb=|\Omega_{M_m}\lb$ as opposed to the massless case.

\subsection{Circle Compactification}

We can equally analyze the cMERA when the scalar field theory is defined on the spacetime cylinder $R\times S^1$
(for the case of $(1+1)$d).
The creation and annihilation operators in the UV theory are denoted by $a^\dagger_{n/R}$ and $a_{n/R}$,
where $n$ is an arbitrary integer and $R$ is the radius of the cylinder.
From the viewpoint of the state $|\Psi(u)\lb$, at the scale $u$ the radius of the cylinder effectively changes into $Re^u$.
Therefore the creation and annihilation operators look like
$a^\dagger_{ne^{-u}/R}$ and $a_{ne^{-u}/R}$. Thus the state $|\Psi(u)\lb$ is simply given by replacing $k$ with $n/R$ in (\ref{mlpsi}).
On the other hand, in the description by the state $|\Phi(u)\lb$, the radius of the cylinder is always $R$ for any $u$.
Thus we just need to replace $k$ with $n/R$ in (\ref{mlphi}) and (\ref{pppf}).

\end{document}